\documentclass[sigconf]{acmart}

\setcopyright{none}
\copyrightyear{© 2026 The Authors}
\acmDOI{}
\acmISBN{}
\acmConference[Accepted at NLBSE '26]{5th International Workshop on Natural Language-based Software Engineering }{April 12--18, 2026}{Rio de Janeiro, Brazil}
\acmBooktitle{5th International Workshop on Natural Language-based Software Engineering (NLBSE '26), April 12--18, 2026, Rio de Janeiro, Brazil}

\usepackage[utf8]{inputenc}
\usepackage[T1]{fontenc}
\usepackage{booktabs}

\usepackage[dvipsnames]{xcolor}
\definecolor{IndexingStageColor}{HTML}{edf5ff}
\definecolor{RetrievalStageColor}{HTML}{fff3d9}
\definecolor{GenerationStageColor}{HTML}{e3fae3}

\begin{document}

\title{From Issues to Insights: RAG-based Explanation Generation from Software Engineering Artifacts}

\author{Daniel P{\"o}ttgen}
\email{danielpoettgen@outlook.de}
\affiliation{%
  \institution{University of Cologne}
  \city{Cologne}
  \country{Germany}
}

\author{Mersedeh Sadeghi}
\email{mersedeh.sadeghi@uni-koeln.de}
\affiliation{%
  \institution{University of Cologne}
  \city{Cologne}
  \country{Germany}
}

\author{Max Unterbusch}
\email{max.unterbusch@uni-due.de}
\affiliation{%
  \department{paluno -- The Ruhr Institute for Software Technology}
  \institution{University of Duisburg-Essen}
  \city{Essen}
  \country{Germany}
}

\author{Andreas Vogelsang}
\email{andreas.vogelsang@uni-due.de}
\affiliation{%
  \department{paluno -- The Ruhr Institute for Software Technology}
  \institution{University of Duisburg-Essen}
  \city{Essen}
  \country{Germany}
}

\renewcommand{\shortauthors}{P{\"o}ttgen et al.}

\begin{abstract}
The increasing complexity of modern software systems has made understanding their behavior increasingly challenging, driving the need for explainability to improve transparency and user trust. Traditional documentation is often outdated or incomplete, making it difficult to derive accurate, context‑specific explanations. Meanwhile, issue‑tracking systems capture rich and continuously updated development knowledge, but their potential for explainability remains untapped.
With this work, we are the first to apply a Retrieval-Augmented Generation (RAG) approach for generating explanations from issue-tracking data.
Our proof-of-concept system is implemented using open-source tools and language models, demonstrating the feasibility of leveraging structured issue data for explanation generation.
Evaluating our approach on an exemplary project's set of GitHub issues, we achieve 90\% alignment with human-written explanations. 
Additionally, our system exhibits strong faithfulness and instruction adherence, ensuring reliable and grounded explanations.
These findings suggest that RAG-based methods can extend explainability beyond black-box ML models to a broader range of software systems, provided that issue-tracking data is available---making system behavior more accessible and interpretable.

\end{abstract}




\maketitle

\section{Introduction}
In recent years, interactions with software systems have become an integral part of everyday life.
As these systems grow in complexity, users increasingly demand transparency to understand system behavior and make informed decisions.
In response, the field of \textit{Explainability} has emerged, aiming to make complex systems more comprehensible~\cite{chazette_exploring_2021, chazette_explainability_2020, sadeghi2024smartex}.
Studies have shown that explanations can increase trust, confidence, and  overall system usability
~\cite{jacovi_formalizing_2021, sadeghi_explaining_2024, eiband_impact_2019}.

Despite this progress, the explainability of general software systems has received limited attention.
Existing solutions are often tailored to AI-based black-box models and are not easily transferable to general software applications.
Furthermore, static explanations (whether manually authored or generated through automated frameworks) are time-consuming to produce and lack adaptability and maintainability.
As systems evolve, such static explanations quickly become outdated, requiring constant maintenance to stay accurate and relevant.

This challenge is further compounded by the difficulty of maintaining complete and accurate documentation in fast-paced, continuously evolving software environments. Although comprehensive documentation is intended to support development processes, it is often deprioritized in practice, resulting in artifacts that are incomplete, outdated, or entirely neglected after initial creation~\cite{mishra_structured_2023, forward_relevance_2002, stettina_necessary_2011, theunissen_mapping_2022, liu_prioritizing_2021}.
In the absence of accurate and up-to-date documentation, generating and validating precise, context-specific explanations becomes increasingly difficult.


To address this gap, we explore issue-tracking systems, such as GitHub Issues and Atlassian Jira, as an alternative and dynamic knowledge source for explanation generation. 
Though initially designed for task management, they have evolved into central communication hubs that support collaboration across agile teams~\cite{bertram_communication_2010}. Unlike static documentation, which is often incomplete or outdated~\cite{mishra_structured_2023, forward_relevance_2002, stettina_necessary_2011, theunissen_mapping_2022, liu_prioritizing_2021}, issue trackers provide continuously updated, fine-grained records of real-time development activity, including design rationales, feature discussions, bug reports, and change justifications. This rich contextual data presents an underexplored opportunity for generating user-facing, natural language explanations of system behavior and evolution.

In this paper, we demonstrate that issue-tracking data can serve as a viable foundation for generating explanations of system behavior. We propose a Retrieval-Augmented Generation (RAG) approach that systematically utilizes issue-tracking data to generate user-facing, natural language explanations. Our proof-of-concept implementation is evaluated on the open-source messaging platform \textit{Mattermost}, showing that accurate and trustworthy explanations can be produced solely from issue-tracking data using compact open-weight Large Language Models (LLMs). To assess its effectiveness, we conduct a systematic evaluation showing that the generated explanations achieve over 90\% alignment with human-written references, while maintaining high faithfulness and instruction adherence. Overall, our findings indicate that issue-driven retrieval can provide a practical and sustainable path toward explainability in general software systems.



\section{Background and Related Work}
\label{sec: Background&RelatedWork}
\textbf{Explainable Systems.} The growing complexity of modern software systems, particularly AI/ML-powered ones, has increased the need for explainability. Many state-of-the-art AI models are perceived as ``black boxes'' due to their opaque decision-making processes ~\cite{confalonieri_historical_2021}, leading to limited user trust and hindered adoption ~\cite{chazette_explainability_2020}. While explainability is critical in sensitive domains like healthcare and security ~\cite{rajpurkar_ai_2022, hamet_artificial_2017}, everyday software systems also benefit from mechanisms that help users understand functionality and improve interactions ~\cite{unterbuschExplanationNeedsApp2023, droste_explanations_2024, sadeghi2021cases, trapp2025facts}. 

An \textit{Explainable System} presents explanations about its functionality, behavior, or decisions to stakeholders, fostering better system understanding and trust ~\cite{chazette_exploring_2021, bussone_role_2015}. While model-specific techniques like SHAP and LIME help interpret ML outputs ~\cite{lundberg_unified_2017, ribeiro_why_2016}, they are not always applicable to everyday software systems where users require more intuitive explanations.

Natural language explanations offer an accessible alternative for both technical and non-technical users ~\cite{cambria_survey_2023}. Recent research has explored automating natural language explanation generation ~\cite{costa_automatic_2018}, with LLMs emerging as powerful tools for this task ~\cite{lubos_llm-generated_2024, pang_generating_2024}. Notable examples include converting model-based explanations to natural language ~\cite{zytek_explingo_2024} and enabling interactive dialogues for ML model explanations ~\cite{slack_explaining_2023}, highlighting LLMs' increasing role in making complex systems more accessible.

\textbf{Large Language Models.} Large Language Models have transformed natural language processing through their ability to understand, analyze, and generate human language ~\cite{wang_pre-trained_2023}. The transformer architecture ~\cite{vaswani_attention_2017} enabled scalable training on extensive datasets, leading to general-purpose LLMs like GPT-4 and Llama 3 that demonstrate unprecedented capabilities in text generation, summarization, and code completion with minimal fine-tuning ~\cite{brown_language_2020, ouyang_training_2022}. 

Current research focuses on improving efficiency, interpretability, and robustness ~\cite{bommasani_opportunities_2022}. Despite their own interpretability challenges, LLMs are increasingly leveraged in \textit{Explainable Systems} to generate natural language explanations ~\cite{lubos_llm-generated_2024, pang_generating_2024}. While domain-specific fine-tuning has proven effective ~\cite{thirunavukarasu_large_2023}, it requires significant computational resources ~\cite{kuang_federatedscope-llm_2024}. This has driven research into few-shot and zero-shot approaches through techniques like prompt engineering ~\cite{kojima_large_2023, sahoo_systematic_2024}. 

The rapid adoption of LLMs raises challenges regarding computational costs, data privacy, and accountability ~\cite{bommasani_opportunities_2022}. The development and training of LLMs demand vast computational resources and extensive datasets ~\cite{lin_data-efficient_2024}. To address these limitations, recent research has explored efficient alternatives such as data pruning and federated learning ~\cite{lin_data-efficient_2024, kuang_federatedscope-llm_2024}. Retrieval Augmented Generation (RAG) offers a promising approach by augmenting LLMs with external knowledge sources during inference, improving performance without requiring fine-tuning. 

\textbf{Retrieval-Augmented Generation.} Large Language Models are constrained by their training data, leading to challenges when addressing queries requiring up-to-date or domain-specific knowledge ~\cite{es_ragas_2023}. LLMs may return generic responses or produce hallucinations when queried about information outside their training scope ~\cite{gao_retrieval-augmented_2024, zhang_sirens_2023}. Addressing these limitations through retraining or fine-tuning poses significant computational challenges ~\cite{kuang_federatedscope-llm_2024}. 

RAG ~\cite{lewis_OG_retrieval-augmented_2020} addresses these limitations by incorporating external knowledge sources during inference. RAG performs a preliminary retrieval step to identify relevant documents from an external database, which are then incorporated into the LLM input to ground responses in retrieved data ~\cite{lewis_OG_retrieval-augmented_2020}. This enables LLMs to generate answers based on up-to-date or private information while reducing hallucinations ~\cite{es_ragas_2023, wang_searching_2024}. 

\begin{figure}
\centering
\includegraphics[width=\columnwidth]{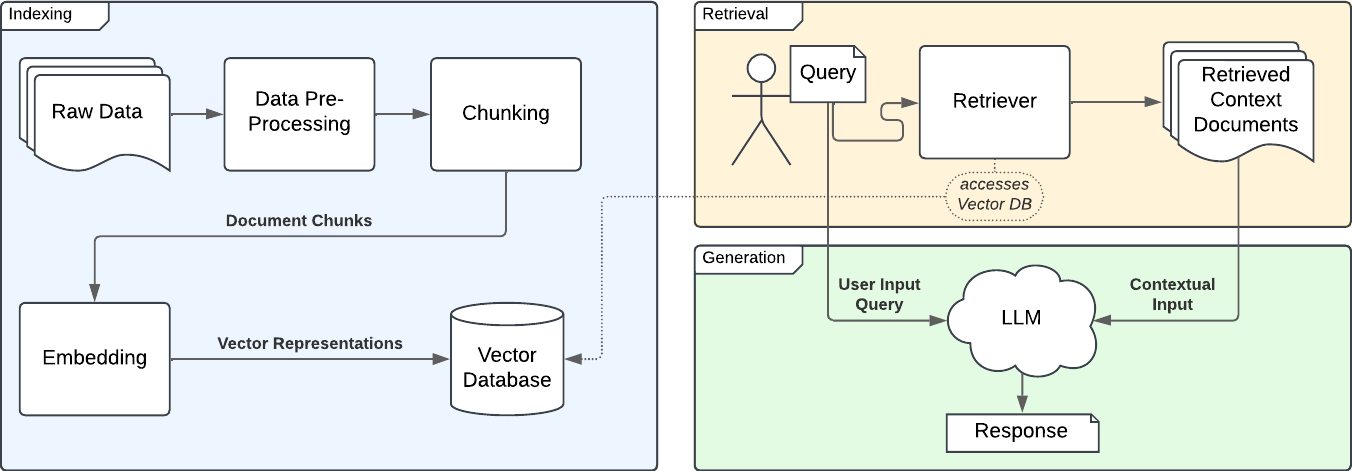}
\caption{RAG Architecture Overview}
\label{fig:architecture}
\end{figure}

A typical RAG pipeline involves indexing documents into a searchable format, retrieving relevant context for a given query, and using this context to generate grounded responses (see Figure \ref{fig:architecture}). Unlike fine-tuning approaches, RAG allows knowledge updates by simply adding new data to the knowledge base without modifying model parameters ~\cite{wang_searching_2024}. Recent research has evolved from simple ``Naive RAG'' approaches to ``Advanced RAG'' techniques that incorporate query optimization, re-ranking, and context refinement to improve retrieval and generation quality ~\cite{gao_retrieval-augmented_2024}. This modularity makes RAG accessible even with API-based LLMs where fine-tuning is not supported ~\cite{es_ragas_2023}. 

\section{Proposed Approach and Implementation}
\label{sec: Approach}
Issue-tracking systems represent a rich but largely underutilized source of knowledge about software behavior. They capture detailed discussions of bugs, feature implementations, design decisions, and user-facing issues—offering contextualized insights into how and why a system behaves as it does. Unlike static documentation, issue data evolves alongside the system, reflecting up-to-date, developer-verified explanations of functionality and changes. Yet, existing explainability approaches rarely leverage this wealth of information, focusing instead on model-centric explanations or static documentation artifacts.

To address this gap, we propose a Retrieval-Augmented Generation approach that transforms issue-tracking data into natural language explanations of system behavior.
The approach provides a foundation for explainability that is inherently data-driven and directly aligned with the software development process.

\subsection{RAG Pipeline Architecture}

Our RAG pipeline consists of three core stages: indexing, retrieval, and generation (see Figure~\ref{fig:architecture}). 
The design emphasizes reproducibility, computational efficiency, and adaptability to different issue-tracking data sources.

\textbf{Indexing.}
Preprocessed issue data is loaded from CSV files and transformed into document objects. We apply recursive chunking with overlap to preserve semantic coherence, and generate embeddings using the \textit{multi-qa-mpnet-base-dot-v1}\footnote{\url{https://huggingface.co/sentence-transformers/multi-qa-mpnet-base-dot-v1}} model from the SentenceTransformers library. This model is specifically optimized for semantic search, allowing it to capture conceptual similarity beyond exact word overlap---an important property for retrieving relevant issue discussions phrased in diverse ways. The resulting embeddings are stored in a persistent Chroma\footnote{\url{https://www.trychroma.com}} vector database for efficient retrieval.

\begin{figure}
\centering
\includegraphics[width=\columnwidth]{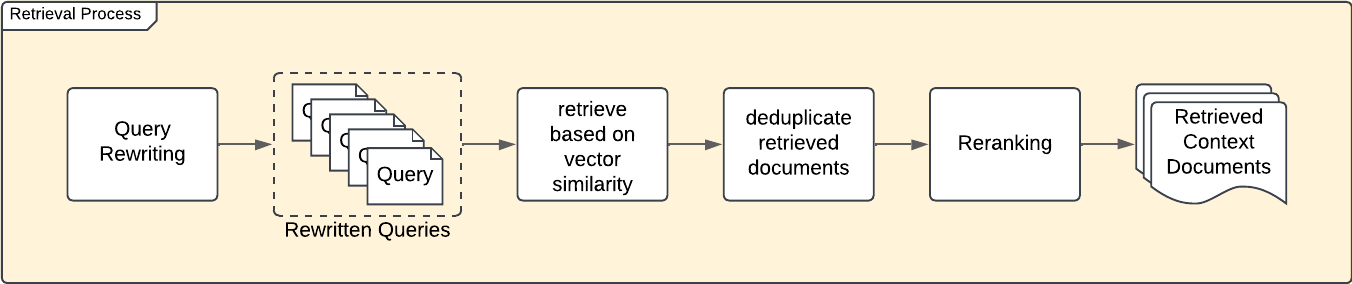}
\caption{Retrieval Process}
\label{fig:Retrieval-Process}
\end{figure}

\textbf{Retrieval.} 
We adopt a multi-stage retrieval process (see Figure~\ref{fig:Retrieval-Process}), that extends beyond the naïve retrieve–read approach, where the top-\(k\) most similar chunks are directly passed to the generator~\cite{gao_retrieval-augmented_2024, ma_query_2023}. Such retrieve-read approaches are sensitive to query-phrasing and retrieval of potentially less relevant context. To address this, we first expand user queries through LLM-based rewriting, which improves robustness against linguistic variation and vague phrasing through alternate query formulations.
Specifically, we utilize the \textit{IBM Granite 3.1 Dense}\footnote{\url{https://ollama.com/library/granite3.1-dense:8b}} model with 8B parameters, selected for its fast inference and strong performance in preliminary tests.
Each rewritten query is matched against the embedded issue corpus via semantic similarity search, leveraging embeddings that capture conceptual rather than purely lexical overlap.

The candidate set of retrieved documents is then post-processed through deduplication and reranking using \textit{Cohere Rerank 3.0}\footnote{\url{https://cohere.com/rerank}}. The model is configured to return 15 chunks, prioritizing semantically central and non-redundant pieces of information, sorted by decreasing relevance.
This ensures that only the most relevant and diverse context is surfaced, reducing the risk of the ``lost-in-the-middle'' effect~\cite{liu_lost_2023} and improving downstream generation quality.

\textbf{Generation.}  
The final context is passed to a locally-deployed LLM through Ollama, combined with a tailored system prompt that constrains the model to rely exclusively on retrieved issue content. The prompt prioritizes faithfulness over fluency, instructing the model to abstain from speculation and explicitly acknowledge missing context when necessary. Generated outputs are structured as JSON objects containing the query, explanation, and supporting evidence, facilitating transparency and downstream evaluation.


\subsection{Implementation}

For our first proof-of-concept implementation, we selected \textit{Mattermost}, an open-source messaging platform. Mattermost provides a large, publicly available, and well-structured issue corpus on GitHub---comprising over 10,000 issues that document feature implementations, bug reports, and design discussions. This makes it an ideal representative case for evaluating the potential of issue-tracking data as an explanatory knowledge base.

Issue data was collected using the GitHub CLI, limited to closed issues to ensure that the underlying functionality had been implemented. We applied preprocessing to remove irrelevant or duplicate entries, preserve informative fields (titles, descriptions, comments, timestamps), and normalize text (e.g., URL replacement and whitespace correction). The resulting dataset provides a structured, semantically rich foundation for retrieval and generation.

To remain within the scope of free and open-source resources, we relied on models available via \textit{Ollama}, running locally on a workstation with an NVIDIA RTX 3070 Ti GPU (8GB VRAM) and 32GB RAM.  
This hardware setup constrained us to models with approximately 14B parameters or fewer.  

We initially screened several candidates and selected the following models for systematic evaluation based on their accuracy and latency on the \textit{System-Specific QA} dataset: \textbf{Granite 3.1 (8B)}, \textbf{Phi-4 (14B)}, \textbf{Qwen 2.5 (14B)}, \textbf{Zephyr (7B)}, and \textbf{DeepSeek-R1 (8B)}.  
Among these, Granite 3.1 was also employed as the evaluator model in our LLM-based metrics due to its strong instruction-following performance and efficient inference.  

We deliberately excluded API-based proprietary models (e.g., GPT-4, Claude) to ensure reproducibility, cost-effectiveness, and independence from external service constraints.  
Importantly, demonstrating the effectiveness of our approach with relatively small open-weight models provides a conservative baseline: since larger proprietary models typically yield higher raw performance, our results suggest that the utility of the proposed pipeline is likely to transfer---and potentially improve---when applied to more powerful LLMs.

To enable interaction and transparency, we developed a Streamlit-based application that allows users to pose natural language queries and view both generated explanations and their retrieved supporting context. This interface provides an intuitive environment for demonstration and human evaluation of the RAG pipeline.

\section{Evaluation}
\label{sec: Evaluation}

Evaluating RAG systems is challenging because generative LLM outputs are non-deterministic, expressed in natural language, and difficult to assess with traditional correctness measures. Nevertheless, structured evaluation is essential to benchmark performance and identify areas for improvement ~\cite{fadnis_inspectorraget_2024}.  

To assess the effectiveness of our approach, we investigate the following research questions:

\begin{itemize}
    \item \textbf{RQ1:} How well does the approach perform in terms of explanation quality compared to human-provided reference answers?
    \item \textbf{RQ2:} How faithful are the generated explanations to the retrieved issue-tracking evidence?
    \item \textbf{RQ3:} How effective is the retrieval component in selecting contextually relevant documents?
\end{itemize}

To evaluate these research questions, we curated three complementary datasets tailored to different aspects of system performance (see Section \ref{subsubsec:TestCases}). The first consists of 10 representative queries about the behavior of the Mattermost platform, with reference answers drawn from official documentation, supporting the evaluation of explanation quality. The second is a set of deliberately out-of-domain queries designed to assess whether the system avoids generating hallucinated content when no relevant context is available, addressing instruction faithfulness. Finally, a randomized robustness dataset was created by mismatching questions and reference answers, allowing us to verify the sensitivity of our evaluation metrics to factual correctness.

Evaluating outputs across these datasets requires metrics that go beyond surface-level similarity. In particular, we must assess factual correctness, contextual grounding, and instruction adherence—all within the open-ended, non-deterministic nature of natural language generation. In the following, we describe our metrics and their selection strategy and rationale.

\subsection{LLM-based Evaluation Setup}
We initially considered token-based metrics such as BLEU, ROUGE, METEOR ~\cite{papineni_bleu_2001, banerjee_meteor_2005, lin_rouge_2004}, embedding-based metrics (BERTScore~\cite{zhang_bertscore_2020} and cosine similarity. However, these proved inadequate for our task as they frequently penalized valid paraphrases and failed to detect subtle 
factual inconsistencies(e.g., ``50MB'' vs. ``500MB'' file limit), resulting in poor 
correlation with human judgment. We therefore opted for LLM-based evaluation.

Recent work shows that LLM-based evaluation, also known as \textit{LLM-as-a-judge}, can approximate expert judgment at scale, making it increasingly popular for RAG benchmarking ~\cite{chiang_can_2023, wang_evaluating_2024, es_ragas_2023}. We employed this approach to assess both answer quality and document relevance. The LLM is instructed to compare each generated response to the corresponding reference answer or retrieval context using structured zero-shot prompts. To ensure transparency, each response includes both a numerical score and a textual justification.


In this direction, we utilized the \textit{IBM Granite 3.1 Dense} model as evaluator due to its high inference speed and evaluation quality in preliminary testing. The temperature was set to 0.3 to reduce creative variability in scoring. Prompts were adapted from evaluation frameworks such as LangSmith and RAGAS ~\cite{es_ragas_2023}, following established best practices for reproducibility. The model was instructed to return output in structured JSON format, enabling automated parsing and detailed scoring analysis.

\subsection{Evaluation Dimensions}\label{sec:evalDimention}
To address our research questions, we operationalized four quality metrics
. As denoted in the previous section, each dimension is evaluated using structured LLM-based prompts, designed to produce both numerical scores and justifications. 


\begin{itemize}
    \item \textbf{Answer vs. Reference (Binary \& Ordinal).}  
    Factual correctness is central to our evaluation. Binary scoring (0/1) enforces strict alignment with the reference, while ordinal scoring (0/5/10) provides a coarser but more nuanced view, distinguishing between fully correct, partially correct, and incorrect answers. 
    Both utilize the same evaluation prompt, with a difference in the possible scores that can be assigned.
    This ensures consistency across evaluations and enables the ordinal metric to serve as a deeper view into the details of the assigned scores, highlighting edge cases where the binary metric would assign a 0 or 1. Together, these metrics balance precision with sensitivity to partial correctness.

    \item \textbf{Answer Helpfulness.}  
    Correctness alone is insufficient if the response is not useful to the user. This binary metric captures whether the answer directly addresses the query in a practical way, reflecting real-world utility.

    \item \textbf{Answer Hallucination / Faithfulness.}  
    Since RAG systems rely on retrieved external knowledge, ensuring that responses remain grounded in the provided context is crucial.
    The Answer Hallucination metric assesses whether the generated response is strictly based on the retrieved context documents or introduces hallucinated information.
    This serves as a key measure of system faithfulness.

    \item \textbf{Document Relevance.}  
    Since generation quality depends on retrieval quality, we separately assess whether the retrieved documents contain information relevant to answering the query. This metric isolates the retrieval stage, enabling more fine-grained analysis of pipeline performance.
\end{itemize}

Together, these metrics provide a multi-faceted view of pipeline quality: from factual alignment and utility (Answer vs. Reference, Helpfulness), to reliability and grounding (Faithfulness), to retrieval effectiveness (Document Relevance). This combination allows us to go beyond surface-level similarity and capture the aspects of explanation quality most relevant to our research questions.  
\subsection{Test Case Design}
\label{subsubsec:TestCases}
We curated three complementary datasets of test cases, each targeting a specific aspect of system performance.
Following established practice in RAG evaluation, all datasets consist of questions paired with baseline answers, enabling consistent measurement with the evaluation dimensions introduced in Section \ref{sec:evalDimention}.

\textbf{Test Case I: System-Specific QA .}
The first dataset\footnote{Find the dataset in our online appendix: \url{https://doi.org/10.6084/m9.figshare.30517931}} contains 10 curated questions about the system behavior of Mattermost.
Both the questions and the corresponding reference answers were derived from the official documentation.  
This dataset directly addresses \textit{RQ 1}, evaluating whether issue tracking data alone provides sufficient context to generate meaningful and accurate explanations.  
By comparing system outputs against documentation-based reference answers, we can systematically assess the explanatory power and limitations of issue-driven retrieval.

\textbf{Test Case II: Out-of-Domain Faithfulness .}
A central requirement of RAG is faithfulness: responses must remain strictly grounded in retrieved context and avoid unsupported claims.  
To assess this property, we constructed a dataset of 10 questions and answers entirely unrelated to Mattermost (see online Appendix).  
Half of the questions concern neighboring domains (e.g., Slack, Microsoft Teams), while the remainder draw on general or historical knowledge unrelated to software explainability.  
Answers were generated using a stand-alone LLM (\textit{Qwen 2.5}) without retrieval, and correctness was manually verified. One exception is the WhatsApp group size question, where the answer reflects an outdated limit (256 vs.\ the current 1024), illustrating the impact of model knowledge cutoffs.

Because the RAG system has no access to relevant supporting context for these questions, a faithful response should explicitly indicate the absence of relevant information rather than attempting to answer.
If the system nevertheless produces an answer, this does not necessarily imply that the content is factually wrong, but rather that the model has failed to adhere to the instruction of grounding responses exclusively in retrieved evidence.  
This dataset therefore provides a rigorous test of faithfulness in the sense of instruction adherence: it benchmarks whether different LLMs resist falling back on pre-trained knowledge when context is unavailable, thereby directly contributing to \textit{RQ 2}.

\textbf{Test Case III: Randomized Answer Robustness.}
Finally, to validate the reliability of our evaluation pipeline itself, we created a robustness dataset derived from the \textit{System-Specific QA Test Case} (see online Appendix).  
Here, question–answer pairs were randomly shuffled once, such that reference answers no longer correspond to their associated questions.  
The system continues to generate answers based on the correct questions, but evaluation is conducted against mismatched references.  

Since these reference answers are systematically incorrect in this context, a well-calibrated evaluation metric should assign low similarity scores.  
If, instead, high scores were produced, this would indicate that the metric fails to critically assess correctness.  
This dataset thus serves as a reliability check: while not directly addressing \textit{RQ 1}, it establishes confidence that the LLM-based ``reference vs. similarity'' metrics provide meaningful signals when applied to the primary evaluation tasks.

\subsection{Results}
\label{subsec:results}

Table~\ref{tab:results} summarizes the experimental results of the RAG pipeline on the four metrics across the tested LLMs and the different datasets.
On the system-specific QA test case, our RAG pipeline consistently produced explanations that closely aligned with human-provided reference answers derived from official documentation.
The strongest model variants achieved high ARS scores ($\geq 0.8$) and faithfulness above 90\%, indicating that generated explanations were both accurate and grounded in retrieved issue-tracking data rather than external pre-trained knowledge.
Document relevance remained consistently high (94–100\%), confirming that the retrieval stage effectively identified the most pertinent context for each query. Helpfulness ratings above 98\% further suggest that the generated explanations would be perceived as practically useful.

\begin{table}
\centering
\caption{Experiment Results}
\small
\begin{tabular}{@{}lcccc@{}}
\toprule
\textbf{LLM} & \textbf{ARS} & \textbf{Faith.} & \textbf{Help.} & \textbf{Doc. Rel.} \\
\midrule
\multicolumn{5}{l}{\textit{System-Specific QA Test Case}}\\
Zephyr (7B)          & 0.68 & 0.86 & 1.00 & 1.00 \\
DeepSeek R1 (8B)    & 0.90 & 0.92 & 0.98 & 0.96 \\
Granite 3.1 (8B)    & 0.82 & 0.96 & 0.98 & 0.96 \\
Phi-4 (14B)         & 0.82 & 1.00 & 1.00 & 0.98 \\
Qwen 2.5 (14B)      & 0.90 & 1.00 & 1.00 & 0.94 \\
\midrule
\multicolumn{5}{l}{\textit{Out-of-Domain Faithfulness Test Case}}\\
Zephyr (7B)          & 0.63 & 0.20 & 0.57 & 0.00 \\
DeepSeek R1 (8B)    & 0.20 & 0.73 & 0.90 & 0.03 \\
Granite 3.1 (8B)    & 0.27 & 0.67 & 0.43 & 0.00 \\
Phi-4 (14B)         & 0.00 & 0.97 & 0.60 & 0.07 \\
Qwen 2.5 (14B)      & 0.00 & 0.80 & 0.53 & 0.03 \\
\midrule
\multicolumn{5}{l}{\textit{Randomized Answer Robustness Test Case}}\\
Zephyr (7B)          & 0.03 &     &     &     \\
DeepSeek R1 (8B)    & 0.23 &     &     &     \\
Granite 3.1 (8B)    & 0.17 &     &     &     \\
Phi-4 (14B)         & 0.03 &     &     &     \\
Qwen 2.5 (14B)      & 0.13 &     &     &     \\
\bottomrule
\end{tabular}

\vspace{0.5ex}
\footnotesize
\textit{Note: ARS = Answer Reference Similarity, Faith. = Answer Faithfulness, \\ Help. = Answer Helpfulness,
Doc. Rel. = Document Relevance}

\label{tab:results}
\end{table}

\textbf{Answering RQ1:} Our findings demonstrate that the proposed RAG-based approach achieves explanation quality comparable to human-provided reference answers, reliably generating accurate, contextually grounded, and trustworthy explanations of system behavior.

To assess how well the approach maintains faithfulness when faced with queries beyond the available issue-tracking data, we evaluated model performance on the Out-of-Domain Faithfulness dataset. The results reveal a correlation of model size on instruction adherence and response reliability. Larger open-weight models such as Phi-4 and Qwen 2.5 consistently recognized when relevant information was unavailable, explicitly acknowledging these limitations rather than fabricating unsupported content. This cautious behavior is reflected in their low ARS scores (0.00) combined with high faithfulness values ($\geq$ 0.80), indicating that the models correctly refrained from generating unverifiable explanations. In contrast, smaller models such as Zephyr and Granite 3.1 frequently relied on pre-trained knowledge, leading to unfaithful or hallucinated outputs, as evidenced by their lower faithfulness scores (0.20--0.67).

\textbf{Answering RQ2:} Our results show that the proposed RAG-based approach achieves high faithfulness in out-of-domain scenarios when paired with larger models, which reliably avoid hallucinations and produce genuine, context-grounded explanations, whereas smaller models tend to generate less trustworthy responses.

To evaluate how effectively the retrieval component selected contextually relevant documents, we examined document relevance scores across both in-domain and out-of-domain test cases. In the system-specific QA dataset, the retriever consistently identified highly pertinent issue-tracking records, achieving document relevance values between 94\% and 100\% across all models. This indicates that the retrieval stage robustly surfaced the most relevant context for each query, directly supporting the generation of accurate and faithful explanations. In contrast, for out-of-domain queries, document relevance appropriately dropped to near zero ($\leq 0.07$), showing that the retriever correctly recognized when no relevant context was available and avoided returning misleading information.

\textbf{Answering RQ3:} These results demonstrate that the retrieval component is highly effective and robust, reliably selecting contextually relevant documents in-domain and correctly withholding context when none is available—thereby ensuring that the generated explanations remain accurate, grounded, and trustworthy.

\textbf{Robustness Check.}
To assess the reliability of the evaluation metrics themselves, we analyzed model performance on the Randomized Answer Robustness dataset, where reference answers were intentionally mismatched with unrelated questions. As expected, this produced very low ARS scores, confirming that the similarity metric correctly penalizes mismatched content.

\section{Discussion}

The issue-driven RAG approach demonstrates that existing development artifacts can serve as a viable and sustainable foundation for explainability in software systems. By drawing on issue-tracking data, it leverages information that naturally evolves alongside the system, reflecting the reasoning and decisions of the development team. This provides an explainability basis that is inherently current, context-aware, and grounded in authentic project knowledge. Moreover, since issue-tracking systems are already integral to most modern software workflows, the approach can be adopted with minimal process overhead, offering a cost-effective and maintainable alternative to existing static explainability efforts.

However, the approach also faces important limitations. Its effectiveness depends heavily on the completeness and quality of the issue-tracking data, which can vary widely across projects and teams. In some cases, discussions may be sparse, overly technical, or fragmented, limiting the ability to derive coherent or user-relevant explanations. Furthermore, not all projects use issue trackers as a central communication or documentation channel, making the approach less suitable for environments where development knowledge is primarily exchanged through other media such as chat platforms, internal wikis, or direct communication.

Another consideration is that issue-tracking systems often contain sensitive or confidential discussions related to internal decisions, vulnerabilities, or proprietary designs. In such contexts, generating user-facing explanations from issue data may not always be appropriate or desirable, as it could expose information not intended for external audiences. Hence, while the approach offers transparency and traceability where openness is valued, it may need careful adaptation in commercial or privacy-sensitive domains.

Overall, our results show that the RAG-based approach reliably produces accurate, context-grounded explanations. High faithfulness and document relevance confirm that outputs are based on retrieved issue data rather than pre-trained knowledge, with larger models yielding more cautious and trustworthy explanations, and smaller models being more prone to hallucination. These findings demonstrate the practical utility of the approach for generating dependable explanations from issue-tracking data.

\subsection{Threats to Validity}

\textbf{Construct Validity.} Our evaluation employed LLM-based metrics to assess explanation quality, which provided better alignment with human judgment than traditional NLP metrics such as token overlap or embedding similarity. While these metrics inherit the biases and context limitations of the evaluation model (Granite 3.1), they offer advantages over human evaluation in terms of consistency, scalability, and reproducibility. Human assessments, while valuable, introduce their own threats to construct validity through subjectivity, inter-rater variability, and dependence on individual expertise and preferences. Our automated approach provided a computationally efficient and reliable approximation that enabled systematic evaluation across a large number of explanations. Nevertheless, complementing automated metrics with targeted human evaluation in future work would strengthen confidence in the validity of our quality assessments, particularly for capturing nuanced aspects of explanation utility in real development contexts.

\textbf{Conclusion Validity.} The non-deterministic nature of LLMs introduces variability in generated responses, which may affect the consistency of our results. We mitigated this threat by conducting multiple runs and reporting the average results across iterations. However, the inherent stochasticity of the models means that individual outputs may vary, also affecting reproducibility.

\textbf{External Validity.} Our evaluation was conducted exclusively on the Mattermost project, which serves as a single case study. While we designed our approach to generalize to any software system with sufficient issue-tracking data, the extent to which our findings transfer to other systems remains an open question. Different software ecosystems may exhibit varying characteristics in their issue-tracking practices, documentation quality, and architectural complexity. Replication studies across multiple projects of varying sizes, domains, and development practices are necessary to establish broader applicability and identify potential boundary conditions of our approach.

\subsection{Future Work}

Our findings open several promising directions for future research. First, while contemporary LLMs with extended context windows raise questions about the necessity of retrieval, RAG remains advantageous by improving transparency, efficiency, and precision in focusing on relevant evidence. Future work should systematically compare end-to-end document prompting against retrieval-based methods in the context of explanation generation, assessing trade-offs between scalability, trustworthiness, and user interpretability.

Second, extending the evaluation beyond Mattermost will be crucial to assess generalizability. Applying our pipeline to other software ecosystems can reveal domain-specific challenges and confirm whether issue-tracking data consistently provides a reliable foundation for explanation generation. This also opens opportunities for studying how variations in issue-tracking practices (e.g., granularity, structure, or completeness of issues) influence explanatory quality.

Third, our results highlight both the feasibility and limitations of relying solely on issue-tracking data. Future research could integrate complementary sources such as documentation, release notes, or developer discussions to expand coverage while maintaining faithfulness. Designing strategies for multi-source integration that preserve transparency and avoid hallucination will be particularly valuable.

Finally, while our proof-of-concept demonstrated strong performance with relatively small open-source models, exploring larger proprietary LLMs and more advanced embedding models may further enhance explanation quality. Such studies should balance improvements in performance with considerations of computational cost, latency, and accessibility.

\section{Conclusion}
\label{sec:Conclusion}

This work demonstrated that issue-tracking data can serve as a viable foundation for generating natural language explanations of software system behavior through a RAG-based approach. Our evaluation on the Mattermost platform showed that explanations derived solely from issue data achieve high alignment with human-written references while maintaining high faithfulness to retrieved evidence. Larger open-weight models produced more trustworthy explanations by reliably acknowledging limitations when information was unavailable, whereas smaller models were more prone to hallucination. While effectiveness depends on issue-tracking quality and privacy considerations may limit applicability in certain contexts, our approach offers a practical and maintainable alternative to static documentation, leveraging development artifacts that naturally evolve alongside the system to make software more transparent and understandable for users.

\bibliographystyle{ACM-Reference-Format}
\bibliography{references}

\end{document}